\documentclass{article}
\pdfoutput=1

\PassOptionsToPackage{numbers, compress, round}{natbib}
\usepackage[final]{neurips_2024_ml4ps}
\usepackage[utf8]{inputenc} % allow utf-8 input
\usepackage[T1]{fontenc}    % use 8-bit T1 fonts
\usepackage{hyperref}       % hyperlinks
\usepackage{url}            % simple URL typesetting
\usepackage{booktabs}       % professional-quality tables
\usepackage{amsfonts}       % blackboard math symbols
\usepackage{nicefrac}       % compact symbols for 1/2, etc.
\usepackage{microtype}      % microtypography
\usepackage{xcolor}         % colors
\usepackage{graphicx}
\title{Bumblebee: Foundation Model \\for Particle Physics Discovery}

\author{%
  Andrew J. Wildridge \\
  Department of Physics and Astronomy\\
  Purdue University\\
  West Lafayette, IN 47907 \\
  \texttt{awildrid@purdue.edu} \\
  \And
  Jack P. Rodgers\\
  Department of Physics and Astronomy\\
  Purdue University\\
  West Lafayette, IN 47907 \\
  \texttt{jprodger@purdue.edu} \\
  \AND
  Ethan M. Colbert \\
  Department of Physics and Astronomy\\
  Purdue University\\
  West Lafayette, IN 47907 \\
  \texttt{colberte@purdue.edu} \\
  \And
  Yao Yao \\
  Department of Physics and Astronomy\\
  Purdue University\\
  West Lafayette, IN 47907 \\
  \texttt{yao317@purdue.edu} \\
  \And
  Andreas W. Jung \\
  Department of Physics and Astronomy\\
  Purdue University\\
  West Lafayette, IN 47907 \\
  \texttt{anjung@purdue.edu} \\
  \And
  Miaoyuan Liu \\
  Department of Physics and Astronomy\\
  Purdue University\\
  West Lafayette, IN 47907 \\
  \texttt{liu3173@purdue.edu} \\
}

\begin{document}

\maketitle

\begin{abstract}
Bumblebee is a foundation model for particle physics discovery, inspired by BERT. By removing positional encodings and embedding particle 4-vectors, Bumblebee captures both generator- and reconstruction-level information while ensuring sequence-order invariance. Pre-trained on a masked task, it improves dileptonic top quark reconstruction resolution by 10-20\% and excels in downstream tasks, including toponium discrimination (AUROC 0.877) and initial state classification (AUROC 0.625). The flexibility of Bumblebee makes it suitable for a wide range of particle physics applications, especially the discovery of new particles.
\end{abstract}

\section{Introduction}
\label{sec:intro}
The intersection of machine learning (ML) and particle physics offers immense potential to improve our understanding of fundamental particles and their interactions~\cite{mlreview}. Foundation models like BERT~\cite{bert} excel in capturing complex relationships and achieving state-of-the-art results, but their inherent sequence sensitivity presents challenges in particle physics, where events, represented by particle 4-vectors, are naturally invariant to order. Additionally, developing an effective pre-training objective that aids in particle discovery remains nontrivial.

We propose \textbf{Bumblebee}, a foundation model inspired by BERT but tailored for particle physics. By removing positional encoding, we ensure sequence-order invariance and modify the embedding to capture particle 4-vectors instead of words, allowing Bumblebee to learn both generator-level (truth) and reconstruction-level (observed) information. Our pre-training objective, akin to BERT’s \textit{Cloze} task~\cite{cloze}, trains Bumblebee to learn the kinematics within the event topology and the transformation between the generator and reconstruction levels. This intricate knowledge of the interactions and event topology required to predict any decay product four momenta enables Bumblebee to perform state-of-the-art regression and classification tasks.

Bumblebee significantly outperforms state-of-the-art methods in dileptonic top quark reconstruction, a challenging task due to the presence of two neutrinos, which only manifest as missing transverse energy (MET) in Large Hadron Collider (LHC) detectors. Moreover, the pre-trained model can be fine-tuned to search for a potential top quark-antiquark (\(\mathrm{t\bar{t}}\)) bound state (toponium) and to enhance the degree of quantum entanglement in pair-produced top quarks at the LHC~\cite{afikentanglement}. This opens new pathways for precision measurements in quantum information science at the highest energies yet achieved.

\section{Related Work}
\label{sec:related_work}
Foundation models have been successfully created for data domains such as images~\cite{visionmae, dino, moco}, text~\cite{bert, gpt3, llama}, and speech~\cite{wav2vec, whisper}. So far, applications of foundation models in particle physics have focused on the reconstruction of particle ``objects" such as jets~\cite{maskedSubjet, omnijetAlpha} and tracks~\cite{bertTrack} and fine-tuning for downstream tasks. Beyond foundation models, the transformer architecture has been successful in particle physics from generating events~\cite{tbqqGeneration, jetGeneration} to achieving state-of-the-art performance in the jet-parton assignment problem~\cite{spanet}. This strongly motivates building a foundation model with the transformer architecture for particle physics discovery.

\section{Bumblebee}
\label{sec:bumblebee}
The Bumblebee model is a transformer-based model~\cite{attention} designed for particle physics. Similar to BERT~\cite{bert} and other foundation models, our framework consists of a pre-training and fine-tuning step. We will be using the event topology of the dileptonic \(\mathrm{t\bar{t}}\) as a case study. Dileptonic \(\mathrm{t\bar{t}}\) means that we have a lepton (antilepton), a b antiquark (quark), and an antineutrino (neutrino) in the final state for each top antiquark (quark).

\subsection{Model architecture}
\label{sec:bumblebee_architecture}
Bumblebee is a multilayer bidirectional transformer encoder based on the original implementation detailed in Ref.~\cite{attention}. Unlike traditional transformer encoders, which rely on positional encodings to handle sequence order, Bumblebee eliminates positional encodings to ensure that they remain invariant to the order in which particles are processed. This architectural decision reflects the physical reality that particles in an event are not ordered in any meaningful way, and thus preserving this permutation invariance is critical for accurately modeling particle physics collisions.

We use \(L\) to denote the number of layers, the hidden size as \(d_{model}\), and the number of heads of self-attention as A. Specifically, the Bumblebee model as reported here is (L=8, \(d_{model}\)=768, A=16, Total Parameters=57M). 

\subsection{Input representation}
\label{sec:bumblebee_input}
Bumblebee takes as input the 4-vectors of particles at both the generator and reconstruction levels, allowing it to learn correlations between partonic truth and reconstructed observables, improving downstream prediction quality. Each 4-vector consists of the transverse momentum \(p_{\mathrm{T}}\), pseudorapidity \(\eta\), azimuthal angle \(\phi\), and mass \(m\) of the particle. Due to color confinement, generator-level quarks form jets~\cite{qcdjetsdisc}. Bumblebee receives b-tag scores to indicate the likelihood of a jet originating from a b quark. Non-jet particles and generator-level b quarks are assigned b-tag scores of 0 and 1, respectively. Neutrinos manifest as MET because of the conservation of momentum, with assigned pseudorapidity and mass of zero, because their z-momentum and energy are unmeasured. These five-dimensional vectors are linearly embedded in the \(d_{model}\)-dimensional space.

\begin{figure}[htb!]
    \centering
    \includegraphics[width=0.75\linewidth]{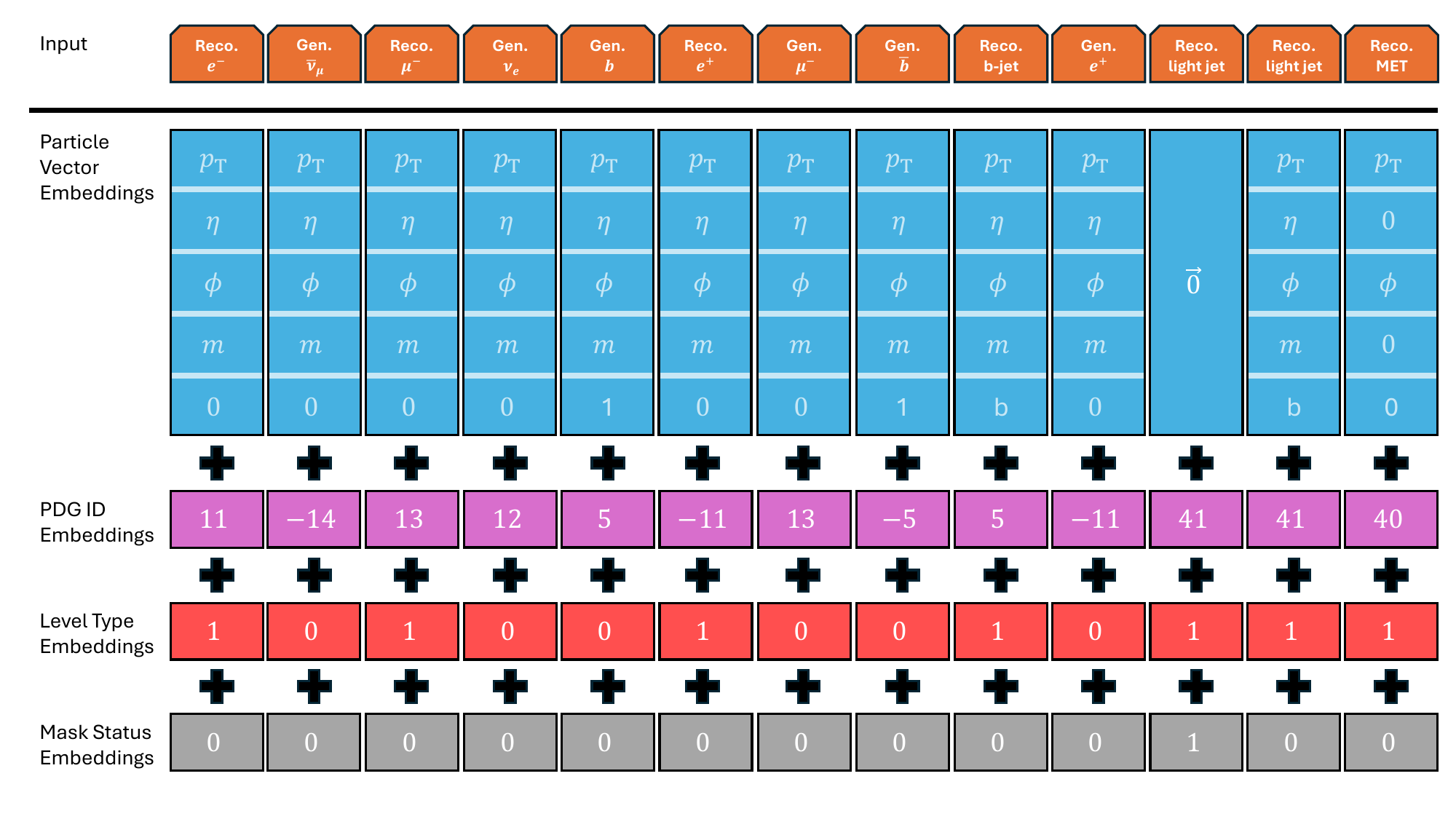}
    \caption{The embedding procedure for Bumblebee for a \(\mathrm{t\bar{t}}\) dileptonic decay. The final embedded input given to Bumblebee is the unweighted sum of the particle vector embedding, PDG ID embedding, level type embedding, and mask status embedding.}
    \label{fig:embedding}
\end{figure}

Additionally, Bumblebee uses three learned embedding tables: (1) to differentiate between reconstruction-level (\texttt{isReco}) and generator-level (\texttt{isNotReco}) particles, (2) to distinguish particle types using a modified PDG ID scheme where b-tagged jets are assigned a PDG ID of 5, non b-tagged jets 41, and MET 40, with all IDs shifted by \(+50\) to map to positive indices, and (3) to indicate whether particles are masked (\texttt{isMasked}) or not (\texttt{isNotMasked}). This masking is essential for Bumblebee's pre-training task (Section~\ref{sec:bumblebee_pretrain}). The final input to Bumblebee is the unweighted sum of the four \(d_{model}\)-dimensional embedded vectors, as illustrated in Fig.~\ref{fig:embedding} for a dileptonic \(\mathrm{t\bar{t}}\) event.

We apply object and event selection criteria typical of a dileptonic \(\mathrm{t\bar{t}}\) analysis at the LHC~\cite{cmsentanglement, cmsdileptonspincorr}. Bumblebee is pre-trained and fine-tuned on events and objects after these selection criteria.

\subsection{Pre-training Bumblebee}
\label{sec:bumblebee_pretrain} 
We pre-train Bumblebee using the \(\mathit{Cloze}\) task~\cite{cloze}, where particles are randomly masked with a \((1/ n_{particles})\%\) probability for half of the training. The other half involves masking all 4-vectors at the generator or reconstruction level. Only the particle vector embedding is masked in both scenarios, shown as \(\vec{0}\) in Fig.~\ref{fig:embedding}. The model minimizes the batch-average mean squared error (MSE) on the predicted masked 4-vectors. Validation and testing of Bumblebee focus on predicting generator-level 4-vectors from reconstruction-level information, framed as dileptonic top quark reconstruction.  During pre-training, we use a linearly decaying learning schedule with a warm-up of 9,000 iterations, peaking at a learning rate of \(\sim10^{-4}\). The Adam optimizer~\cite{adam} is used with \(\beta_1 = 0.9\), \(\beta_2 = 0.999\), and weight decay for regularization. Epsilon, dropout, batch size and weight decay are set to \(10^{-8}\), 0.05, 16, and \(10^{-3}\), respectively. Training is carried out on 2 V100 GPUs for 10 epochs, and the model with the lowest validation loss is used for test predictions.

\subsection{Fine-tuning}
After pre-training Bumblebee on the \(\mathrm{t\bar{t}}\) system, we fine-tune the model for downstream tasks. For classification, a masked vector \((1, 0, 0, 0, 0)\) is added for signal and \((0, 0, 0, 0, 0)\) for background, with a ``PDG ID" of 50, using the reconstruction-level (\texttt{isReco}) and masked (\texttt{isMasked}) embeddings. Generator-level information is omitted during fine-tuning, and the model is trained to predict this vector, similar to the \(\texttt{CLS}\) token in BERT. To avoid catastrophic forgetting~\cite{catforget}, the learning rate is reduced by an order of magnitude. Hyperparameters such as weight decay, dropout, batch size, and \(\epsilon\) are optimized during this phase and trained for 4 epochs.

\section{Datasets}
\label{sec:bumblebee_dataset}
To test the performance of Bumblebee, we generate a 7M \(\mathrm{t\bar{t}}\) Monte Carlo sample at next-to-leading order using \(\textsc{powheg}\textsc{v2}\) ~\cite{Frixione:2007nw, Frixione:2007vw, Nason:2004rx, Alioli:2010xd}. Additionally, we use a toy model~\cite{Fuks_2021} to generate a 1M Monte Carlo sample of the ground state, \(\eta_{\mathrm{t}}\), of the \(\mathrm{t\bar{t}}\) bound state toponium at leading order with \(\textsc{MadGraph5}\_\textsc{aMC@NLO}\)~\cite{mad5}. Both processes perform parton showering and hadronization with \(\textsc{Pythia}\)~\cite{pythia}. Detector simulation is performed with \(\textsc{Delphes}\)~\cite{delphes} using the default card for the CMS detector card. We use a 70/15/15 training/validation/test split for each dataset and all results are shown on the withheld test set.

\section{Experiments}
We conducted several experiments to evaluate the ability of Bumblebee to learn foundational \(\mathrm{t\bar{t}}\) physics and fine-tune on downstream tasks.

\begin{figure}[!htb]
    \centering

    \includegraphics[width=0.375\linewidth]{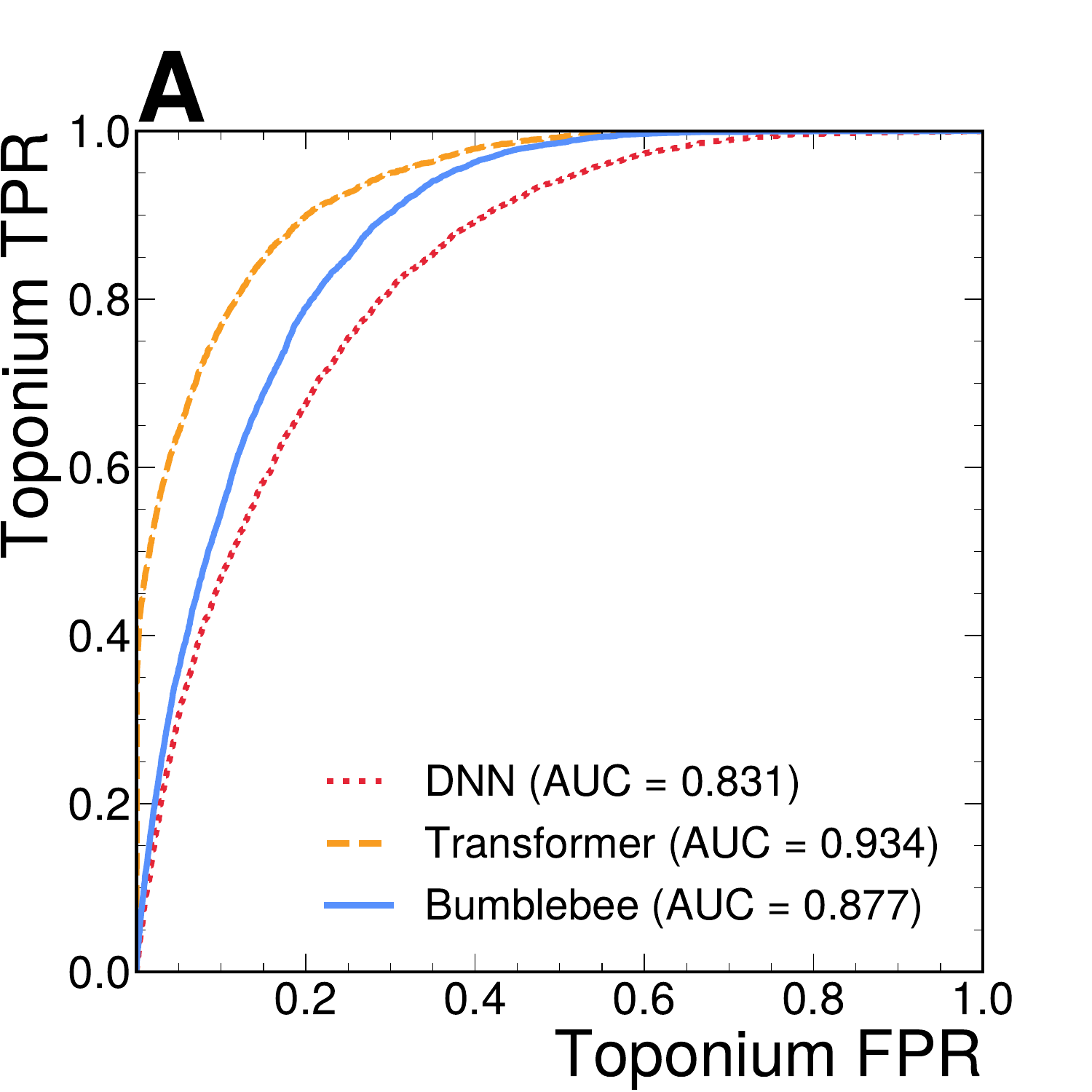}
    \includegraphics[width=0.375\linewidth]{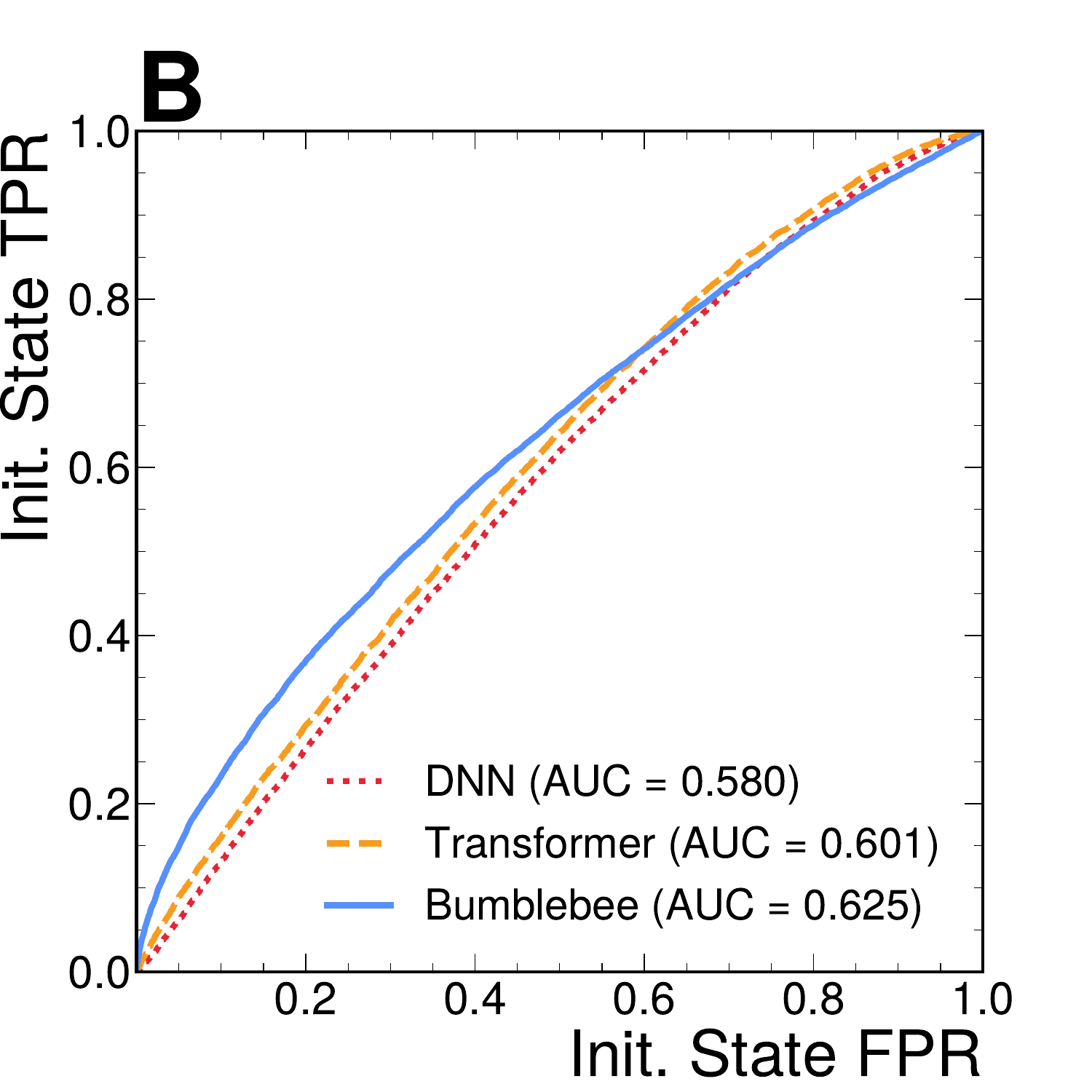} \\
    \includegraphics[width=0.375\linewidth]{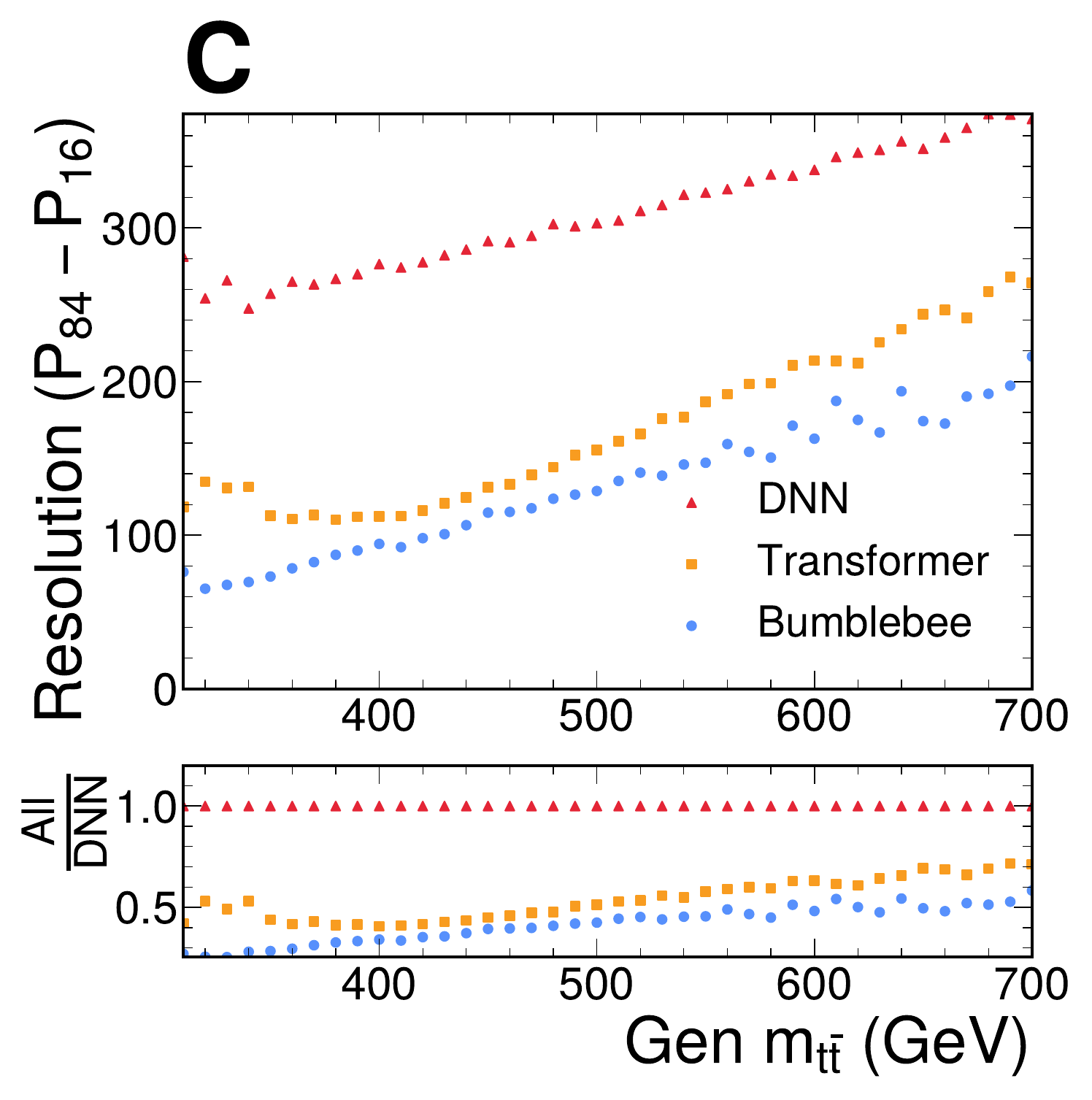}
    \caption{ 
    A) The receiver operating characteristic (ROC) curve for Bumblebee fine-tuned on discriminating toponium against \(\mathrm{t\bar{t}}\). Two supervised models, DNN and Transformer, are shown for comparison.
    B) The ROC curve for Bumblebee fine-tuned on discriminating the initial state of \(\mathrm{t\bar{t}}\). The positive class is the gluon-gluon initial state. Two supervised models, DNN and Transformer, are shown for comparison.
    C) The resolution of \(m(\mathrm{t\bar{t}})\) given as the difference between the 16$^{th}$ and 84$^{th}$ percentiles of the $m(\mathrm{t\bar{t}})$ residuals ($P_{84} - P_{16}$) as a function of the true \(m(\mathrm{t\bar{t}})\).} 
    \label{fig:finetune_results}
\end{figure}

\paragraph{Dileptonic top quark reconstruction}
Our pre-training task doubles as a dileptonic top quark reconstruction challenge. The reconstructed top quark is the sum of predicted generator-level daughter four-vectors. A 10-20\% improvement in the resolution of the \(\mathrm{t\bar{t}}\) system’s invariant mass (\(m(\mathrm{t\bar{t}})\)) is achieved compared
to a supervised transformer, 
%to the state-of-the-art \(m_{\script lb}\) weighting method~\cite{cmsdilepton2015, cmsdilepton2019},
as shown in Fig.~\ref{fig:finetune_results}C. The improved \(m(\mathrm{t\bar{t}})\) reconstruction resolution at high invariant mass is of great importance for heavy resonance searches of physics beyond the Standard Model~\cite{ttbar_resonance}. 

\paragraph{Toponium discrimination}
We benchmark the ability of Bumblebee to discover new particles using the ground state of toponium (\(\eta_{\mathrm{t}}\)), a hypothetical particle predicted by the Standard Model~\cite{Fuks_2021, toponium1, toponium2, toponium3, toponium4}. Due to the low resolution in \(m(\mathrm{t\bar{t}})\) relative to \(\eta_{\mathrm{t}}\)’s width (3 GeV)~\cite{toponiumhunter}, observation of \(\eta_{\mathrm{t}}\) is challenging and an appropriate benchmark for Bumblebee. We fine-tune Bumblebee on weighted binary cross-entropy. With early stopping and a class imbalance of 10:1, the model achieves an AUC of 0.877, as seen in Fig.~\ref{fig:finetune_results}A.

\paragraph{Initial state classification}
Top quark pairs at the LHC originate from gluon-gluon or quark-antiquark interactions with a rich dependence on this origination~\cite{afikentanglement, cmsacmeas}. Initial-state discrimination enhances the search for new physics~\cite{jachargeasym}. Bumblebee achieves a 0.625 AUC in this task, marking the first attempt at initial-state classification and outperforming supervised machine learning models, as shown in Fig.~\ref{fig:finetune_results}B.

\section{Limitations}
\label{sec:limitations}
The primary limitation of this work is its focus on the dileptonic decays of top quark pairs at the LHC, driven by the computational cost of generating Monte Carlo for each process. Although photons are absent in dileptonic decays, there is nothing in our embedding procedure that inherently restricts Bumblebee from handling photons at the reconstruction or generator level. Another limitation is the event topology: in dileptonic decays, the main challenge is not jet-parton assignment but reconstructing the missing neutrino 3-vectors from MET. More complex topologies, such as \(\mathrm{t\bar{t}H}\)~\cite{cmstth, atlastth} and \(\mathrm{t\bar{t}t\bar{t}}\)~\cite{ttttcms, ttttatlas}, feature numerous jets and multiple neutrinos, offering further avenues for exploration.

\section{Ablation study}
%In this section, we present ablation studies for Bumblebee and their importance on the results achieved in the pre-training and fine-tuning tasks.

\begin{figure}[!htb]
    \centering
 
    \includegraphics[width=0.375\linewidth]{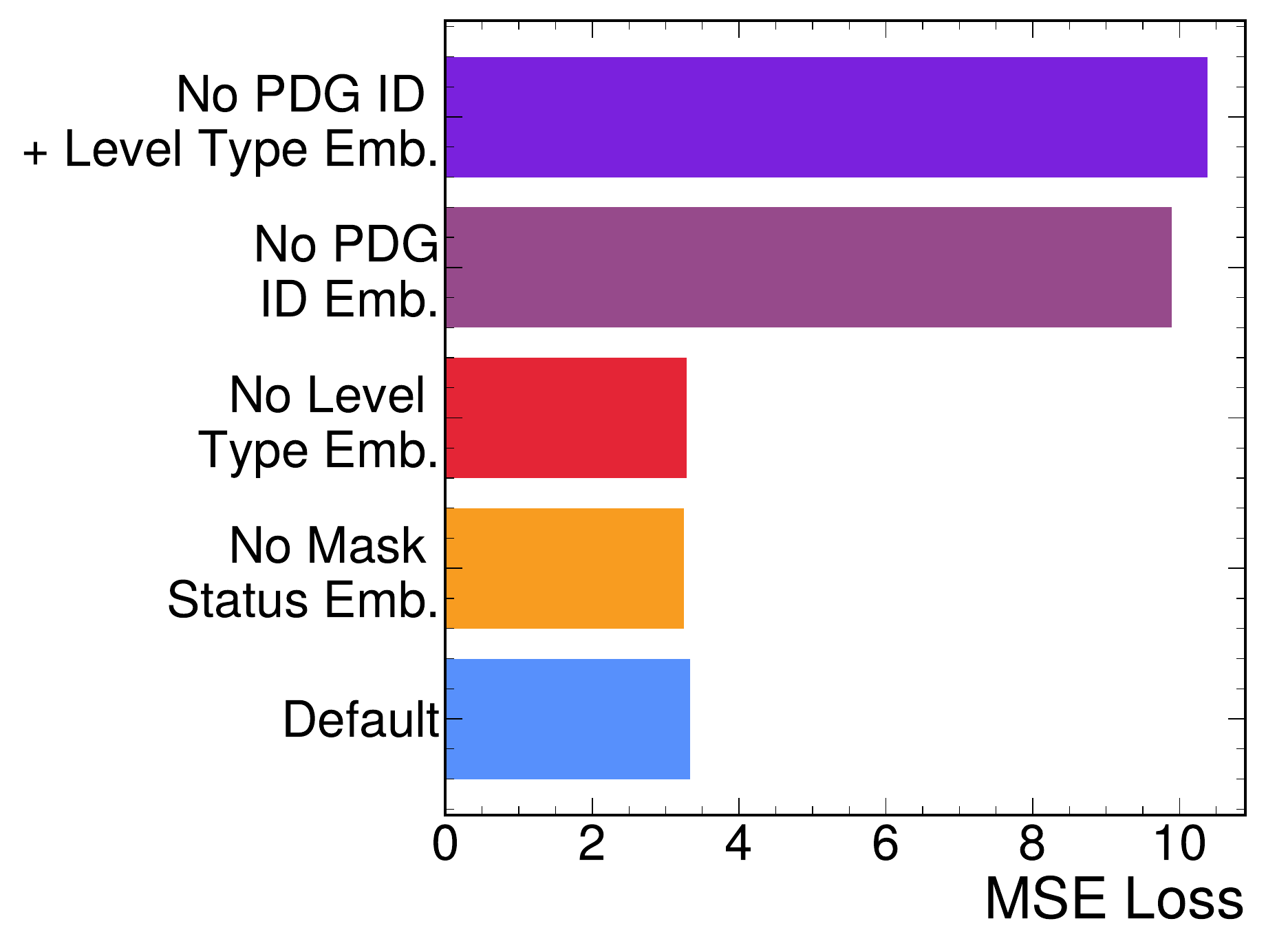}
    \caption{ 
    %A)
    The comparison of the best validation loss obtained on the pre-training objective when removing individual embeddings present in the input representation. The embeddings removed are labeled on the y-axis.
    %B) A comparison of the resolution of \(m(\mathrm{t\bar{t}})\) for the various ablated models.
    } 
    \label{fig:ablation_study}
\end{figure}

%\subsection{Effect of Embeddings}
We present an ablation study where we remove embeddings in the input representation and measure the performance regarding the MSE loss on the validation set. In Fig.~\ref{fig:ablation_study}, it is clear that the most important embedding is the PDG ID embedding as this results in the largest increase of MSE loss. This is expected as this defines the event topology at the generator and reconstruction levels. The next most important embedding is the level type embedding which is obvious when considering scenarios where a particle is masked both at the reconstruction and generator level. 

\section{Conclusion}
The Bumblebee model has demonstrated its ability to outperform state-of-the-art methods in dileptonic top quark reconstruction, while also generalizing this knowledge to various downstream tasks. Its flexibility may extend beyond the \(\mathrm{t\bar{t}}\) process, as any particle with kinematics can be embedded, making it highly versatile for a wide range of applications. Although we focused on specific discrimination and pre-training tasks, Bumblebee can be fine-tuned for other uses. With its strong performance and adaptability, Bumblebee shows evidence of being a possible path to constructing foundation models applicable to many particle physics processes.

\begin{ack}
This material is based upon work supported by the U.S. Department of Energy program under Award Number(s) DE-SC00023700 and AI for a more precise future of the top quark. The authors declare no competing interests. AW thanks M.W. Kerrigan for insightful discussions regarding foundation models and A. Anuar for providing help in producing \(\eta_{\mathrm{t}}\) Monte Carlo.
\end{ack}

\bibliography{neurips_2024.bib}

%\printbibliography

%%%%%%%%%%%%%%%%%%%%%%%%%%%%%%%%%%%%%%%%%%%%%%%%%%%%%%%%%%%%

% \appendix

% \section{Appendix / supplemental material}

% Optionally include supplemental material (complete proofs, additional experiments and plots) in appendix.
% All such materials \textbf{SHOULD be included in the main submission.}

%%%%%%%%%%%%%%%%%%%%%%%%%%%%%%%%%%%%%%%%%%%%%%%%%%%%%%%%%%%%

\end{document}